\documentclass[prl,aps,twocolumn,showpacs,superscriptaddress]{revtex4}%
\usepackage{amsmath}
\usepackage{amsfonts}
\usepackage{amssymb}
\usepackage{graphicx}%

\begin{document}
\title{Non-exponential London penetration depth in Ba$_{1-x}$K$_{x}$Fe$_{2}$As$_{2}$ single crystals}
\author{C.~Martin}
\affiliation{Ames Laboratory and Department of Physics \& Astronomy, Iowa State University, Ames, IA 50011}
\author{R.~T.~Gordon}
\affiliation{Ames Laboratory and Department of Physics \& Astronomy, Iowa State University, Ames, IA 50011}
\author{M.~A.~Tanatar}
\affiliation{Ames Laboratory and Department of Physics \& Astronomy, Iowa State University, Ames, IA 50011}
\author{H.~Kim}
\affiliation{Ames Laboratory and Department of Physics \& Astronomy, Iowa State University, Ames, IA 50011}
\author{N.~Ni}
\affiliation{Ames Laboratory and Department of Physics \& Astronomy, Iowa State University, Ames, IA 50011}
\author{S.~L.~Bud'ko}
\affiliation{Ames Laboratory and Department of Physics \& Astronomy, Iowa State University, Ames, IA 50011}
\author{P.~C.~Canfield}
\affiliation{Ames Laboratory and Department of Physics \& Astronomy, Iowa State University, Ames, IA 50011}
\author{H.~Luo}
\affiliation{Institute of Physics, Chinese Academy of Sciences, Beijing 100190, China}
\author{H.~H.~Wen}
\affiliation{Institute of Physics, Chinese Academy of Sciences, Beijing 100190, China}
\author{Z.~Wang}
\affiliation{Institute of Physics, Chinese Academy of Sciences, Beijing 100190, China}
\author{A.~B.~Vorontsov}
\affiliation{Department of Physics, Montana State University, P.O. Box 173840}
\author{V.~G.~Kogan}
\affiliation{Ames Laboratory and Department of Physics \& Astronomy, Iowa State University, Ames, IA 50011}
\author{R.~Prozorov}
\email[Corresponding author: ]{prozorov@ameslab.gov}
\affiliation{Ames Laboratory and Department of Physics \& Astronomy, Iowa State University, Ames, IA 50011}

\date{10 February 2009}

\begin{abstract}
We have studied the in- and out-of-plane magnetic penetration depths in the hole- doped iron based superconductor Ba$_{1-x}$K$_{x}$Fe$_{2}$As$_{2}$ ($T_{c}\approx$ 30K). The study was performed on single crystals grown from different fluxes and we find that the results are nearly the same. The in-plane London penetration depth $\lambda_{ab}$ does not show exponential saturation at low temperature, as would be expected from a fully gapped superconductor. Instead, $\lambda_{ab}(T)$ shows a power-law behavior, $\lambda\propto T^{n}$ ($n\approx 2$), down to $T\approx 0.02~T_{c}$, similar to the electron doped Ba(Fe$_{1-x}$Co$_{x}$)$_{2}$As$_{2}$. The penetration depth anisotropy $\gamma_{\lambda}=\lambda_{c}(T)/\lambda_{ab}(T)$ increases upon cooling, opposite to the trend observed in the anisotropy of the upper critical field, $\gamma_{\xi}=H_{c2}^{\perp c}(0)/H_{c2}^{\parallel c}(0)$. These are universal characteristics of both the electron and hole doped 122 systems, suggesting unconventional superconductivity. The behavior of the in-plane superfluid density $\rho_{ab}(T)$ is discussed in light of existing theoretical models proposed for the iron pnictides superconductors.
\end{abstract}

\pacs{74.25.Nf,74.20.Rp,74.20.Mn}
\maketitle

The discovery of superconductivity ($T_{c}\simeq26$~K) in fluorine doped LaFeAsO ("1111")~\cite{Kamihara08} has generated remarkable interest in the community. In a short time, the critical temperature was increased by pressure or chemical substitution above 55 K, which is significantly larger than the highest $T_c$ reported in any s-wave superconductor, i.e. MgB$_{2}$, and comparable to those of the cuprates. Later, superconductivity with $T_{c}$ as high as 38 K was discovered in Ba$_{1-x}$K$_{x}$Fe$_{2}$As$_{2}$~\cite{Rotter08}. This BaK-122 compound is particularly important because, unlike the cuprates or the 1111  iron pnictides, it is not an oxide, downplaying the role of oxygen in this type of high temperature superconductors. Moreover, large, high quality single crystals of both electron and hole doped 122 pnictides were synthesized~\cite{Sefat08, Ni08_Co, Ni08_K, Luo08}, which is essential for drawing reliable conclusions regarding their physical properties. 

One key feature for understanding the origin of the high critical temperature and the pairing mechanism in pnictide superconductors is the symmetry of the order parameter. Phase diagrams of electron~\cite{Ni08_Co} and hole-doped~\cite{Rotter08_2} 122 show that superconductivity emerges through doping, suppressing the orthorhombic/antiferromagnetic (AF) ground state in the parent compound. Close proximity to a magnetic state could imply the importance of magnetic fluctuations for pairing and may be reflected in the symmetry of the superconducting gap. At the same time, band structure calculations and ARPES experiments~\cite{Cvetkovic08, Liu08} show that multiple bands cross the Fermi level, opening the possibility for multiband superconductivity. Several ARPES studies~\cite{Ding08, Zhao08, Nakayama08} on Ba$_{1-x}$K$_{x}$Fe$_{2}$As$_{2}$ have found at least two different superconducting gaps without nodes in the $ab$-plane. Point contact spectroscopy~\cite{Szabo08}, specific heat~\cite{Mu08} and microwave penetration depth~\cite{Hashimoto08} data also suggest fully gapped superconductivity, and possibly two gaps. However, $\mu$-SR~\cite{Goko08} studies show a linear temperature dependence of the superfluid density at low temperatures and spin-lattice relaxation rate 1/T$_1$, from $^{75}$As NMR~\cite{Fukazawa09} was found to be proportional to $T^3$. Together with more recent reversible magnetization measurements~\cite{Salem09}, these results leave open the possibility for a nodal gap. Extensive work on the electron doped 122 via Co substitution (FeCo-122)~\cite{Gordon08_1, Gordon08_2} have revealed that at low temperatures the penetration depth does not show exponential saturation, but instead it features a robust power-law behavior $\Delta\lambda(T)\propto T^{n}$, with $n$ being between 2 and 2.5, depending on the doping level.  

In this paper, we report measurements of the London penetration depth in the hole doped 122 compound Ba$_{1-x}$K$_{x}$Fe$_{2}$As$_{2}$. To account for the effect of sample preparation on experimental data and draw objective conclusions, single crystals of Ba$_{1-x}$K$_{x}$Fe$_{2}$As$_{2}$ with $T_{c}\approx$ 30 K, grown by two different groups, using different fluxes, were studied. A total of six samples was measured, giving highly reproducible results. We show the data for two of them. The crystal labeled $A$, with dimensions of 740$\times$800$\times$70$\mu$m$^3$ was grown out of Sn flux. It comes from one of the batches characterized in Ref.~\cite{Ni08_K}. Based on elemental analysis from Ref.~\cite{Ni08_K}, sample A corresponds to optimal K doping with $x$=0.45, i.e. Ba$_{0.55}$K$_{0.45}$Fe$_{2}$As$_{2}$. Sample $B$, with dimensions of 320$\times$830$\times$70~$\mu$m$^3$, was grown from self-flux of FeAs and it was characterized in Ref.~\cite{Luo08}. According to Ref.~\cite{Luo08}, sample B corresponds to $x$=0.3, i.e. Ba$_{0.7}$K$_{0.3}$Fe$_{2}$As$_{2}$.

The penetration depth was measured using a tunnel diode resonator (TDR) technique~\cite{Degrift74}, by inserting the sample into the inductor of a self-resonating tank circuit powered by a tunnel diode. The resonant frequency of the empty resonator was $f_{0}\approx 14~MHz$, with a stability better than 5 ppb/hour. The rf magnetic field produced by the inductor was H$_{rf}\sim 10$\,mOe, much less than typical values for the lower critical field $H_{c1}\sim 50$\,Oe in iron-arsenides. Therefore, when a superconducting sample is placed inside the inductor, the inductance of the coil changes due to Meissner screening, which leads to a change in the resonant frequency. The relative frequency change is directly proportional to the penetration depth, $\Delta f=-G\Delta\lambda$~\cite{Prozorov00}. The calibration constant $G$ was determined both by following a calibration procedure described in Refs.~\cite{Prozorov00,Prozorov06} and from in-situ extraction of the sample from the coil, with both methods yielding identical results. First, we mounted the sample with the c-axis along the rf magnetic field ($H_{rf}\parallel c$). In this geometry, only the in-plane screening currents were induced, and therefore $\Delta f\propto \Delta\lambda_{ab}$. Then, the sample was aligned with $H_{rf}\perp c$. For this geometry, screening currents flow both in the $ab$ plane and along the $c$ direction and $\Delta\lambda_{c}$ could be obtained in our relatively thick samples using the numerical model described in Ref.~\cite{Prozorov06}. 

\begin{figure}[tb]
\includegraphics[width=.85\columnwidth]{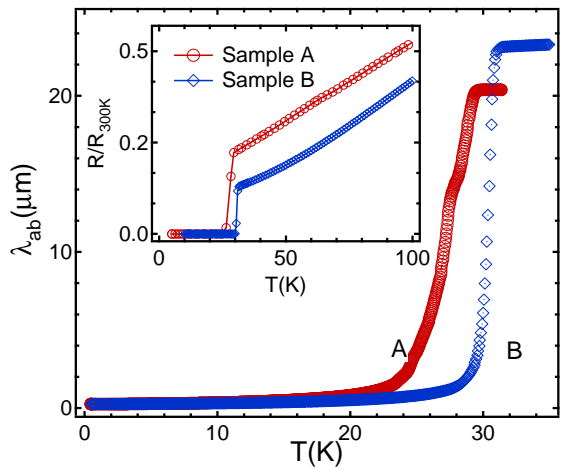}
\caption{(Color online) $\lambda_{ab}(T)$ for sample A, grown from Sn flux~\cite{Ni08_K} and for sample B, grown from FeAs self-flux~\cite{Luo08}. Inset: Temperature dependence of resistivity for each sample, normalized to the room temperature value.}
\label{Fig1}
\end{figure}

Figure~\ref{Fig1} shows $\lambda_{ab}(T)$ (main figure) and resistivity (inset) for samples A and B. We notice that although they correspond to different potassium concentrations, both samples have very similar values for the onset $T_c$. The widths of the transition at $T_c$ are significantly different, with $\Delta T\leq$ 3K for sample A and $\Delta T\leq$ 1.5K for sample B. Moreover, despite the fact that optimal doping corresponds to $x=0.45$ K for both methods, the maximum $T_c$ reached by using Sn flux is about 30 K~\cite{Ni08_K}, whereas by using FeAs flux it is 38 K~\cite{Luo08}. A possible explanation is that some amount of Sn is incorporated in the crystal structure of sample A, as explained in Ref.~\cite{Ni08_K}.

\begin{figure}[tb]
\includegraphics[width=.85\columnwidth]{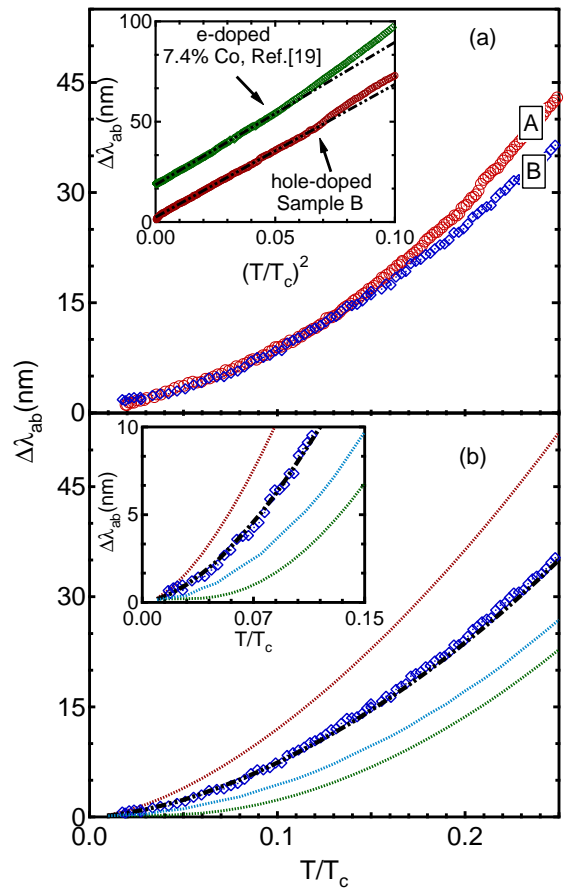}
\caption{(Color online) (a) Low temperature region of $\lambda_{ab}(T)$ for Sample A and Sample B, respectively. Inset: $\Delta\lambda_{ab}$ vs. (T/T$_c$)$^2$ for hole doped (Sample B   from present work) and electron doped ($x$=0.074 Co from Ref.~\cite{Gordon08_1}) FeAs-122 (shifted vertically for clarity).  Dashed lines represent linear fits. (b) $\lambda_{ab}(T)$ for Sample B and our theoretical calculations based on the model from Ref.~\cite{Vorontsov09}, for the following values of impurity scattering parameters ($\Gamma_{0}$, $\Gamma_{\pi}$) (top to bottom): (0,0.064), (3, 0.064), (3,0.068), (3,0.060). Inset: The lowest temperature region, showing good agreement between experiment and theory.} 
\label{Fig2}
\end{figure}

Figure~\ref{Fig2} shows the low temperature region of $\lambda_{ab}(T)$ for samples A and B. Despite the differences in potassium content and transition width at $T_{c}$, the penetration depth behavior is similar below 0.3$T_c$, exhibiting a power law $\lambda_{ab}(T)=\lambda_{0}+bT^n$, with the average value of the exponent being $n\approx 1.9\pm 0.4$. It is important to note that this nearly quadratic temperature dependence of the penetration depth is similar to previous observations in the electron-doped 122~\cite{Gordon08_1, Gordon08_2}. 

The inset to Fig.~\ref{Fig2} shows $\Delta\lambda_{ab}$ plotted against ($T/T_c$)$^2$ for sample B and for a Ba(Fe$_{1-x}$Co$_{x}$)$_{2}$As$_{2}$ single crystal ($x$=0.074) \cite{Gordon08_1}, corresponding to optimal doping. It can be clearly observed that at low temperatures, both traces are well fit by straight lines with similar slopes. This very similar power-law behavior suggests that there are common electrodynamic properties within the 122 pnictide superconductors.

There are several possible explanations for this observed behavior. The existence of point nodes on 3D parts of the Fermi surface was suggested in Refs.~\cite{Gordon08_1,Gordon08_2}. A quadratic temperature dependence for the penetration depth is also expected for a superconducting gap with line nodes in the presence of strong (unitary) impurities, which will create an additional quasiparticle density of states~\cite{Hirschfeld93}. It has also been proposed that antiferromagnetic spin fluctuations could mediate {\it inter-band} pairing in the iron pnictides. In this case, the order parameter is fully gapped but changes sign between different Fermi surface sheets.  This situation is referred to as the extended  $s^{+}$ symmetry ~\cite{Mazin08}. It was found that within this approach either the existence of line nodes between the two Fermi surfaces~\cite{Parish08} and/or impurity scattering~\cite{Vorontsov09} can explain a power-law temperature dependence of the experimentally observed NMR relaxation rate and penetration depth. In Fig~\ref{Fig2}(b) we plot the experimental data for sample B together with our calculations considering the extended s$^+$ symmetry ($\Delta\propto$cos$k_{x}$+cos$k_{y}$) in the presence of non-magnetic impurities for several values of the intra-band ($\Gamma_{0}$) and inter-band ($\Gamma_{\pi}$) scattering rates~\cite{Vorontsov09}. Details of these calculations are given in Ref.~\cite{Vorontsov09}, where this model was used for the Co doped 122 system. With $\lambda_{ab}(0)$ as a fitting parameter, we were able to correctly reproduce the experimental data down to the lowest measured temperature, as seen in the inset of Fig.~\ref{Fig2}(b).

\begin{figure}[tb]
\includegraphics[width=.85\columnwidth]{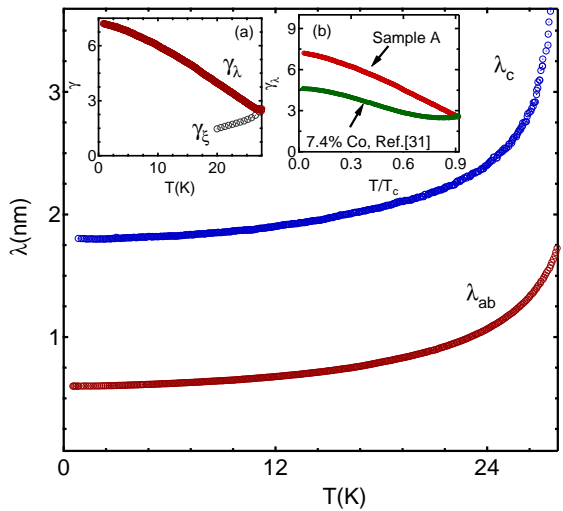}
\caption{(Color online) In-plane $\lambda_{ab}$(T) and out-of-plane $\lambda_{c}$(T) penetration depth for Sample B. Inset (a): Variation with temperature of $\gamma_{\lambda}=\lambda_{c}/\lambda_{ab}$ obtained by dividing the values from the main panel, and $\gamma_{\xi}(T)$ from Ref.~\cite{Altarawneh08}. Inset (b): $\gamma_{\lambda}$ for hole (current work) and electron doped (Re.~\cite{Tanatar08}) FeAs-122.} 
\label{Fig3}
\end{figure}

However, the fit yields $\lambda_{ab}(0)\approx$ 600 nm while experimentally it was found to be about 180 nm~\cite{Li08, Prozorov09}. Further theoretical and experimental investigations may reconcile this discrepancy and it is possible that the power-law dependence of $\lambda(T)$ results from the effect of strong inter-band scattering on the order parameter with extended s-wave symmetry. However, the possibility for an anisotropic or even nodal superconducting gap remains open, particularly given the recent report of linear behavior of $\lambda$(T) in LaFePO crystals~\cite{Fletcher08}. We only conclude that our results for the hole doped Ba$_{1-x}$K$_{x}$Fe$_{2}$As$_{2}$ suggest unconventional superconductivity in pnictide superconductors.

Applying the rf magnetic field both parallel to the $c$-axis and along the $ab$-plane, we were able to directly subtract the change in penetration depth associated with screening currents flowing in the $c$-direction, i.e. $\Delta\lambda_{c}$. Since we are not aware of independent estimates of $\lambda_{c}$ at any temperature, we have used the approach of Ref.~\cite{Tanatar08}. High magnetic field experiments on single crystals of Ba$_{1-x}$K$_{x}$Fe$_{2}$As$_{2}$~\cite{Ni08_K, Altarawneh08}  found that $\gamma_{\xi}=H_{c2}^{ab}(0)/H_{c2}^{c}(0)\approx 3.5$ near $T_c$. Based on the validity of the Ginzburg-Landau theory in the vicinity of $T_c$, we consider $\gamma_{\xi}=\gamma_{\lambda}\approx 3.5$ at $T=0.9~T_{c}$. Figure~\ref{Fig3} shows the resulting $\lambda_{c}(T)$ on the same graph with $\lambda_{ab}(T)$. Because $\lambda_{c}$ has a weaker temperature dependence at low temperatures, it extrapolates to a significantly higher value at $T$=0, $\lambda_{c}(0)\approx$ 1800 nm. Therefore, we obtain a penetration depth anisotropy $\gamma_{\lambda}\approx 7$ near $T=0$, which is about twice the value of $\gamma_{\lambda}(T_{c})$. In inset (b) of Fig.~\ref{Fig3}, we show $\gamma_{\lambda}(T)$ from the present work together with the result for the electron doped Ba(Fe$_{0.926}$Co$_{0.074}$)$_{2}$As$_{2}$ from Ref.~\cite{Tanatar08}. In both cases, $\gamma_{\lambda}$ decreases with temperature, although at different rates, which emphasizes again the similarity between the hole and electron doped 122 superconductors.

We now compare the temperature dependence of the penetration depth anisotropy $\gamma_{\lambda}(T)$ with that of the upper critical field
$\gamma_{\xi}(T)$. In pnictide superconductors, $H_{c2}$(0) may reach $\sim100~T$, making it difficult to estimate $\gamma_{\xi}$ over the entire temperature range~\cite{Altarawneh08}. Nevertheless, it has been found that $\gamma_{\xi}$ decreases from $\approx 3.5$ near $T_c$ to $\approx 1.4$ at 0.5T$_{c}$~\cite{Altarawneh08} and we display both $\gamma_{\xi}(T)$ and $\gamma_{\lambda}(T)$ in inset (a) of Fig.~\ref{Fig3}. A possible explanation for the difference in the temperature dependence of the anisotropies could be given by the two-gap scenario. In a clean superconductor at $T=0$, $\gamma_{\lambda}^{2}(0)=\langle v_a^2\rangle/\langle v_c^2\rangle$, where $v_{a}$ and $v_{c}$ are the Fermi velocities along the a and c directions. The averaging is performed over the whole Fermi surface. However, at $T=T_c$, $\gamma_{\lambda}^{2}(T_{c})=\langle\Delta^{2}v_a^2\rangle/\langle\Delta^{2}v_c^2\rangle$, where $\Delta$ is the superconducting gap. For two gaps with different values, the main contribution to $\gamma_{\lambda}(T_{c})$ is given by the regions of the Fermi surface with a larger gap. If the larger gap exists on a less anisotropic Fermi surface, it will result in  $\gamma_{\lambda}(T_{c})<\gamma_{\lambda}(0)$. This situation is the reverse to that of MgB$_2$, where $\gamma_{\lambda}$ increases with temperature~\cite{Fletcher05}. However, unlike MgB$_2$, as we show below, a simple two gap s-wave  model cannot correctly describe the superfluid density,  $\rho_{ab}(T)=\left(\lambda_{ab}(0)/\lambda_{ab}(T)\right)^{2}$ in  Ba$_{1-x}$K$_{x}$Fe$_{2}$As$_{2}$.

\begin{figure}[tb]
\includegraphics[width=.9\columnwidth]{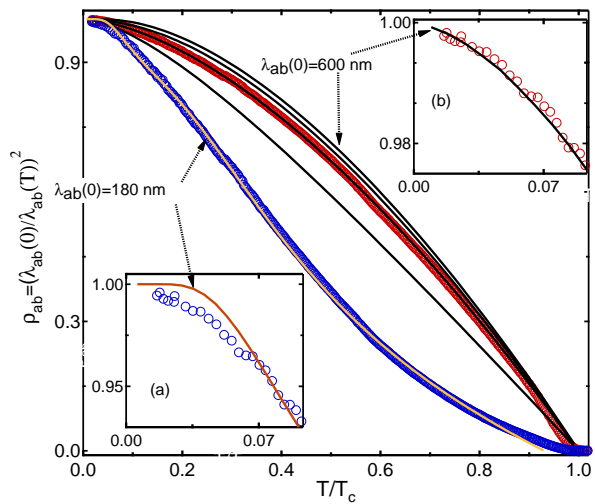}
\caption{(Color online) Blue symbols: $\rho_{ab}(T)$ for $\lambda_{ab}(0) = 180~nm$. Orange line shows best fit to the two-gap s-wave model. Red symbols: $\rho_{ab}(T)$ for $\lambda_{ab}(0) = 600~nm$. Black lines show calculated $\rho_{ab}(T)$ for the same values of the scattering parameters, $\Gamma_{0}/2\pi~T_{c0}$ and $\Gamma_{\pi}/2\pi~T_{c0}$, shown in Fig.~\ref{Fig2}, obtained from the fit to extended s-wave model. 
Inset (a): Zoom of $\rho_{ab}(T)$ for $\lambda_{ab}(0) = 180~nm$ (symbols) with the best fit to two-gap model. Inset (b): Zoom of $\rho_{ab}(T)$ for $\lambda_{ab}(0) = 600~nm$ (symbols) with the best fit to $s^{+}$ model.} 
\label{Fig4}
\end{figure}

Figure~\ref{Fig4} shows $\rho_{ab}(T)$ for Sample B calculated for $\lambda_{ab}(0)$=180 nm and 600 nm. For the experimental value of $\lambda_{ab}(0)$=180 nm (blue circles), $\rho_{ab}(T)$ has positive curvature at $T>T_c/2$, whereas both s- and d-wave calculations predict only negative curvature in the entire temperature range. This may be explained assuming two independent s-wave gaps $\Delta_{1}(T)$ and $\Delta_{2}(T)$, with relative contributions $\epsilon$ and $(1-\epsilon)$, so that $\rho(T)$=$\epsilon\rho(\Delta_{1})$+ $(1-\epsilon)\rho(\Delta_{2})$. Orange line shows the best fit to this model. Even though the quality of the fit looks acceptable at intermediate temperatures, the model clearly fails to reproduce the low temperature data, as illustrated in the inset (a) of Fig.~\ref{Fig4}. Moreover, the parameters obtained from the fit: $\Delta_{1}(0)\approx 0.95 k_{B}T_{c}$, $\Delta_{2}(0)\approx 0.25 k_{B}T_{c}$ and $\epsilon\approx~0.8$ are physically unreasonable. Both gaps are smaller than the BCS weak coupling limit of $1.76~k_{B}T_c$ and they are much smaller than those previously reported from ARPES~\cite{Ding08, Zhao08}. These two gaps would reproduce measured T$_c$ only if inter-band coupling is strong, in which case our fitting model of independent gaps is not applicable ~\cite{Nicol05}. We have tried the same fit for $\lambda_{ab}(0)$=600 nm and arrived at the same conclusion. Furthermore, theoretical calculations using the extended $s^{+}$ symmetry in the presence of impurities do not reproduce the experimental data for $\lambda_{ab}(0)$=180 nm either. On the other hand, if we use $\rho_{ab}(T)$ calculated with $\lambda_{ab}(0)$ = 600 nm (red circles in Fig.~\ref{Fig4}) a good agreement with $s^{+}$ model can be achieved in the entire temperature range. Black lines in Fig.~\ref{Fig4} show $\rho_{ab}(T)$ calculated for four sets of scattering parameters, $\Gamma_{0}/2\pi~T_{c0}$ and $\Gamma_{\pi}/2\pi~T_{c0}$, shown in Fig.~\ref{Fig2}. Inset (b) zooms on the low-temperature part and indicates that the best fit to $s^{+}$ model works down to the lowest temperature.

In conclusion, the hole doped compound, Ba$_{1-x}$K$_{x}$Fe$_{2}$As$_{2}$, exhibits a nearly quadratic temperature variation of the London penetration depth down to 0.02T$_{c}$, at odds with the exponential behavior expected for conventional fully gapped superconductor. This is similar to the behavior observed for electron doped Ba(Fe$_{1-x}$Co$_{x}$)$_{2}$As$_{2}$ \cite{Gordon08_1,Gordon08_2} and suggests unconventional superconductivity in the entire 122 pnictide family. As a note, we mention that similar measurements of rf penetration depth by the Bristol group also found $\lambda_{ab}(T)\propto T^{2}$ at low temperatures~\cite{Carrington}.

We thank A.~V.~Chubukov, I.~I.~Mazin, J.~Schmallian and M.~G.~Vavilov for stimulating discussions and A.~Carrington for discussions and sharing unpublished data. Work at the Ames Laboratory was supported by the Department of Energy-Basic Energy Sciences under Contract No. DE-AC02-07CH11358. M.A.T. acknowledges continuing cross-appointment with the Institute of Surface Chemistry, NAS Ukraine. R. P. acknowledges support from Alfred
P. Sloan Foundation.


\begin{thebibliography}{99}                                                                                               %

\bibitem {Kamihara08}Y.~Kamihara \emph{et al.}, J. Am. Chem. Soc. \textbf{130} 3296 (2008).

\bibitem {Rotter08}M.~Rotter \emph{et al.}, Phys. Rev. B \textbf{78}, 020503 (2008). 

\bibitem{Sefat08}A.~S.~Sefat \emph{et al.}, Phys. Rev. Lett.  \textbf{101}, 177004 (2008).
 
\bibitem {Ni08_Co}N.~Ni \emph{et al.}, Phys. Rev. B \textbf{78}, 214515 (2008).  

\bibitem {Ni08_K}N.~Ni \emph{et al.}, Phys. Rev. B \textbf{78}, 014507 (2008). 

\bibitem{Luo08}H.~Luo \emph{et al.}, Supercond. Sci. Technol. \textbf{21}, 125014 (2008) 

\bibitem {Rotter08_2}M.~Rotter \emph{et al.},Chem. Int. Ed. \textbf{47} (2008). 

\bibitem{Cvetkovic08}V.~Cvetkovic \emph{et al.}, arXiv.org:0804.4678 (2008).

\bibitem {Liu08}C.~Liu \emph{et al.}, Phys. Rev. Lett. \textbf{101}, 177005 (2008). 

\bibitem {Ding08}H.~Ding \emph{et al.}, Europhys. Lett., \textbf{83}, 47001 (2008). 

\bibitem {Zhao08}L.~Zhao \emph{et al.}, Chin. Phys. Lett. \textbf{25}, 4402-4405 (2008). 

\bibitem{Nakayama08}N.~Nakayama \emph{et al.}, arXiv.org:0812.0663 (2008). 

\bibitem{Szabo08}P.~Szabo \emph{et al.}, arXiv.org:0809.1566 (2008).

\bibitem{Mu08}G.~Mu  \emph{et al.}, arXiv.org:0812.1188 (2008). 

\bibitem{Hashimoto08}K.~Hashimoto \emph{et al.}, arXiv.org:0810.3506 (2008).

\bibitem {Goko08}T.~Goko \emph{et al.}, arXiv.org:0808.1425 (2008). 

\bibitem{Fukazawa09}H.~Fukazawa \emph{et al.}, arXiv.org:0901.0177 (2009). 

\bibitem{Salem09}S.~Salem-Sugui Jr., \emph{et al.}, arXiv.org:0902.1252 (2009).

\bibitem{Gordon08_1}R.~T.~Gordon \emph{et al.}, arXiv.org:0810.2295 (2008).

\bibitem{Gordon08_2}R.~T.~Gordon \emph{et al.}, arXiv.org:0812.3683 (2008).

\bibitem{Degrift74}C.~T.~Van Degrift, Rev. Sci. Instrum. \textbf{46}, 599 (1974)

\bibitem{Prozorov00}R. Prozorov \textit{et al.}, Phys. Rev. B \textbf{62}, 115 (2000)

\bibitem{Prozorov06}R. Prozorov \textit{et al.}, Supercond. Sci. Techn. \textbf{19}, R41 (2006)

\bibitem{Hirschfeld93}P.~J.~Hirschfeld \textit{et al.}, Phys. Rev. B \textbf{48}, 4219 (1993)

\bibitem {Mazin08}I.~I.~Mazin \emph{et al.}, Phys. Rev. Lett. \textbf{101}, 057003 (2008). 

\bibitem{Parish08}M.~M.~Parish \textit{et al.}, Phys. Rev. B \textbf{78}, 144514 (2008)

\bibitem{Vorontsov09}A.~B.~Vorontsov \textit{et al.}, arXiv.org:0901.0719 (2009).

\bibitem{Li08}G.~Li \emph{et al.}, Phys. Rev. Lett. \textbf{101}, 107004 (2008) 

\bibitem{Prozorov09}R. Prozorov \textit{et al.},arXiv.org:0901.3698 (2009).

\bibitem{Fletcher08}J.~D.~Fletcher \emph{et al.}, arXiv.org:0812.3858 (2008).

\bibitem{Tanatar08}M.~A.~Tanatar  \emph{et al.}, arXiv.org:0812.4991 (2008).

\bibitem{Altarawneh08}M.~M.~Altarawneh \textit{et al.}, Phys. Rev. B \textbf{78}, 220505 (2008).

\bibitem{Fletcher05}J.~D.~Fletcher \emph{et al.}, Phys. Rev. Lett. \textbf{95}, 097005 (2005) 

\bibitem{Nicol05}E.~J.~Nicol \textit{et al.}, Phys. Rev. B \textbf{71}, 054501 (2005).

\bibitem{Carrington}A.~Carrington, Private Communication

\end{thebibliography}
\end{document}